\documentclass[reprint,twocolumn,superscriptaddress,pre,showpacs,amsmath,amssymb,aps]{revtex4-1}

\usepackage{natbib}	
\usepackage{graphicx,color,rotating}
\usepackage[latin1]{inputenc}
\usepackage{textcomp}
\usepackage{dcolumn}% Align table columns on decimal point
\usepackage{bm}     % bold math
\usepackage{upgreek}
\usepackage{subdepth}  			% Added by JTK (aligns sub-indices)
\usepackage{multirow}
\usepackage{hyperref}					% Package for hyperreferences
\usepackage{xcolor}						% Allows for coloring of text
\hypersetup{			     			% Setup of the Hyperref package
    colorlinks,
    linkcolor={blue!80!black},
    citecolor={blue!80!black},
    urlcolor={blue!80!black}
}
\usepackage{array}
\newcolumntype{L}[1]{>{\raggedright\let\newline\\\arraybackslash\hspace{0pt}}m{#1}}
\newcolumntype{C}[1]{>{\centering\let\newline\\\arraybackslash\hspace{0pt}}m{#1}}
\newcolumntype{R}[1]{>{\raggedleft\let\newline\\\arraybackslash\hspace{0pt}}m{#1}}
\newcommand{\mr}[1]{\ensuremath{\mathrm{#1}}}
\renewcommand{\vec}[1]{\bm{#1}}
\newcommand{\ee}{\mathrm{e}}
\newcommand{\ii}{\mathrm{i}}
\newcommand{\dm}{\mathrm{d}}

\newcommand{\avr}[1]{\big\langle #1 \big\rangle}

\newcommand{\iot}{{\ii\omega t}}

\newcommand{\pp}{\partial}

\newcommand{\nablabf}{\boldsymbol{\nabla}}

\newcommand{\divop}{\nablabf\cdot}

\newcommand{\eee}{\vec{e}}

\newcommand{\fffac}{\vec{f}_\mathrm{ac}}

\newcommand{\gvec}{\vec{g}}

\newcommand{\rrr}{\vec{r}}

\newcommand{\vvv}{\vec{v}}

\newcommand{\zerovec}{\boldsymbol{0}}

\newcommand{\cO}{c_0}

\newcommand{\Eac}{E_\mathrm{ac}}

\newcommand{\kapO}{\kappa_0}

\newcommand{\pI}{p_1}

\newcommand{\vvvI}{\vvv_1}

\newcommand{\vvvII}{\vvv_2}

\newcommand{\rhoO}{\rho_0}

\newcommand{\rhoI}{\rho_1}

\newcommand{\SICel}{^\circ\!\textrm{C}}

\newcommand{\SImuL}{\textrm{\textmu{}L}}

\newcommand{\SIkgm}{\textrm{kg}\:\textrm{m$^{-3}$}}

\newcommand{\SIm}{\textrm{m}}

\newcommand{\SImum}{\textrm{\textmu{}m}}

\newcommand{\SIPa}{\textrm{Pa}}

\newcommand{\SIs}{\textrm{s}}
\newcommand{\SIms}{\textrm{ms}}

\newcommand{\sigmabf}{\bm{\sigma}}

\newcommand{\sigmabfI}{\bm{\sigma}^{{}}_1}

\newcommand{\eqlab}[1]{\label{eq:#1}}
\renewcommand{\eqref}[1]{Eq.~(\ref{eq:#1})}

\newcommand{\eqsref}[2]{Eqs.~(\ref{eq:#1}) and~(\ref{eq:#2})}

\newcommand{\figref}[1]{Fig.~\ref{fig:#1}}

\newcommand{\figlab}[1]{\label{fig:#1}}

\newcommand{\secref}[1]{Section~\ref{sec:#1}}

\newcommand{\seclab}[1]{\label{sec:#1}}
\newcommand{\tabref}[1]{Table~\ref{tab:#1}}

\newcommand{\tablab}[1]{\label{tab:#1}}

\begin{document}
%\preprint{Preprint identifier}

\title{Characterization of Acoustic Streaming in Gradients of Density and Compressibility}

\author{Wei Qiu}
\email{weiqiu@fysik.dtu.dk}
\affiliation{Department of Physics, Technical University of Denmark, DTU Physics Building 309, DK-2800 Kongens Lyngby, Denmark}

\author{Jonas T. Karlsen}
\affiliation{Department of Physics, Technical University of Denmark, DTU Physics Building 309, DK-2800 Kongens Lyngby, Denmark}

\author{Henrik Bruus}
\email{bruus@fysik.dtu.dk}
\affiliation{Department of Physics, Technical University of Denmark, DTU Physics Building 309, DK-2800 Kongens Lyngby, Denmark}

\author{Per Augustsson}
\email{per.augustsson@bme.lth.se}
\affiliation{Department of Biomedical Engineering, Lund University, Ole R\"{o}mers v\"{a}g 3, 22363, Lund, Sweden}

\date{16 October 2018}

\begin{abstract}
Suppression of boundary-driven Rayleigh streaming has recently been demonstrated for fluids of spatial inhomogeneity in density and compressibility owing to the competition between the boundary-layer-induced streaming stress and the inhomogeneity-induced acoustic body force. Here we characterize acoustic streaming by general defocusing particle tracking inside a half-wavelength acoustic resonator filled with two miscible aqueous solutions of different density and speed of sound controlled by the mass fraction of solute molecules. We follow the temporal evolution of the system as the solute molecules  become homogenized by diffusion and advection. Acoustic streaming rolls is suppressed in the bulk of the microchannel for 70--200 seconds dependent on the choice of inhomogeneous solutions. From confocal measurements of the concentration field of fluorescently labelled Ficoll solute molecules, we conclude that the temporal evolution of the acoustic streaming depends on the diffusivity and the initial distribution of these molecules. Suppression and deformation of the streaming rolls are observed for inhomogeneities in the solute mass fraction down to $0.1~\%$.
\end{abstract}

\maketitle

% Main text

\section{Introduction}
\seclab{Intro}

Acoustic streaming is a steady flow that arises in a fluid medium interacting with sound waves. It has been studied extensively~\cite{Nyborg1953a, Nyborg1958, Lighthill1978, Riley2001, Muller2015} because of its important role in thermoacoustics~\cite{Bailliet2001, Hamilton2003a}, medical ultrasound~\cite{Marmottant2003, vanderSluis2007, Wu2008a, Doinikov2010}, and acoustic levitation~\cite{Trinh1994, Yarin1999}. Acoustic streaming has been classified into two categories based on its formation mechanisms. One mechanism is the spatial attenuation of acoustic waves in the bulk of the fluid, which results in a time-averaged net force in the direction of the wave propagation~\cite{Eckart1948, Nyborg1953a}. This type of streaming, called quartz wind or bulk-driven Eckart streaming, is generally observed in large systems where the length scale of wave propagation is much longer than the wavelength. The other mechanism, predominant in systems of a size comparable to the wavelength, such as the system under investigation in this work, is that of acoustic energy dissipation in the viscous boundary layers, where the velocity of the oscillating fluid decays to match the velocity of the boundary~\cite{LordRayleigh1884, Schlichting1932} of either walls~\cite{Nyborg1958, Hamilton2003, Muller2013} or suspended objects~\cite{Elder1959, Lee1990, Yarin1999, Tho2007}. This  boundary-driven so-called Rayleigh streaming typically generates a recirculating flow in the bulk.

Rayleigh streaming has been identified as a key limiting factor in standing-wave, acoustic particle manipulation~\cite{Bruus2011c, Barnkob2012a, Hammarstrom2012, Collins2015, Marin2015, Hahn2015a, Guo2016} because suspended microparticles are subject to both acoustic radiation forces and Stokes drag forces from the acoustic streaming. The relative magnitude of the two forces depends on the microparticle size and the material properties of the particle and the suspending fluid. For microparticles below a critical size, the motion of microparticles is dominated by acoustic streaming, which in many cases hinders the manipulation of sub-micrometer sized particles. Manipulation below the classical limit has previously been demonstrated by flow vortices generated by two-dimensional acoustic fields~\cite{Antfolk2014, Mao2017}, by acoustically active seed particles~\cite{Hammarstrom2012}, by a thin reflector design~\cite{Carugo2014}, or in systems actuated by surface acoustic waves~\cite{Collins2017, Wu2017, Sehgal2017}.

Recently, we discovered that an acoustic body force can cause relocation and stabilization of inhomogeneities in fluids of spatially inhomogeneous density and compressibility when subjected to a standing wave field~\cite{Deshmukh2014, Karlsen2016, Karlsen2017}. This spurred the development of iso-acoustic focusing, an equilibrium method to measure cell acoustic properties wherein cells migrate in a fluid of gradually increasing acoustic impedance to their points of zero acoustic contrast~\cite{Augustsson2016}. Furthermore, the acoustic body force caused by a spatial inhomogeneity in density was found to enable efficient suppression of acoustic streaming in the bulk inside a half-wavelength resonator ~\cite{Karlsen2018}. This finding paves the way for acoustic manipulation, fractionation, and in-line sample preparation of sub-micrometer particles of biological relevance such as bacteria, virus and exosomes, as well as trapping of hot plasma in gasses \cite{Koulakis2018}

Here, we extend the study of acoustic streaming to fluids made inhomogeneous in both density and speed of sound by the addition of different solute molecules, and we investigate its evolution in a ultrasound half-wavelength glass-silicon resonator with rectangular cross-section. The suppression of acoustic streaming is mapped for different combinations of gradients in density and speed of sound. The evolution of the acoustic streaming and the molecular concentration field is measured in fluids of different solute molecule concentration and diffusivity by particle tracking velocimetry and confocal microscopy, respectively. We conclude that the acoustic streaming is strongly dependent on inhomogeneities in the solute mass fraction down to $0.1~\%$.

\section{Theory of inhomogeneous acoustofluidics}
\seclab{Theory}

We briefly summarize the theory of acoustic streaming and its suppression in inhomogeneous fluids based on our full account in Ref.~\cite{Karlsen2018}. We consider a fluid that is made inhomogeneous by adding solute molecules with dilute-limit diffusivity $D$ and a spatiotemporal dependent mass ratio (concentration) $s = s(\rrr,\tau)$. The physical properties of the resulting solution thus depends on space and time through $s$:  Density $\rho_0(s)$, sound speed $c_0(s)$, compressibility $\kappa_0(s) = (\rho_0c_0^2)^{-1}$,  and viscosity $\eta_0(s)$. Moreover, the solute molecules have an $s$-dependent diffusivity $D(s)$. As discussed in Refs.~\cite{Karlsen2016, Karlsen2018}, a crucial property of this system, when placed in an acoustic field, is the separation of time scales between the fast acoustics $t\sim 0.1~\upmu$s and the slow hydrodynamics $\tau\sim 10~\SIms$. Because $\tau \sim 10^5 t$, the acoustic fields can be computed while keeping the hydrodynamic degrees of freedom fixed at each instance in time $\tau$.

\subsection{Fast-time-scale acoustics}
The inhomogeneous solutions is placed an acoustic cavity where an time-harmonic standing acoustic wave is imposed at  frequency $f$ and angular frequency $\omega = 2\pi f$. We assume the usual adiabatic case for the first-order pressure field $p_1$, density field $\rhoI$, and velocity field $\bm{v}_1$ of amplitude $p_\mathrm{ac}$, $\rho_\mathrm{ac}$, and $v_\mathrm{ac}$, respectively. Writing each acoustic field on the form $\rho = \rhoO(\rrr,\tau) + \rhoI(\rrr,\tau)\:\ee^{-\iot}$, the governing equations become~\cite{Karlsen2018},
 \begin{subequations}
 \eqlab{FirstOrderEqs}
 \begin{align}
 - \ii \omega \rhoO \vvvI &= \divop\sigmabfI ,\\
 - \ii \omega \kapO \pI &= - \divop\vvvI , \\
 - \ii \omega \rhoO \kapO \pI &= - \ii \omega \rhoI + \vvvI \cdot \nablabf \rhoO .
 \end{align}
 \end{subequations}

\subsection{The acoustic body force}

As we have shown in Ref.~\cite{Karlsen2016}, the acoustic fields acting on the short time scale $t$ give rise to an acoustic body force $\bm{f}_\mathrm{ac}$ acting on the inhomogeneous fluid on the slow time scale $\tau$. This body force is derived from the nonzero divergence in the time-averaged (over one oscillation period $2\pi/\omega$) acoustic momentum-flux-density tensor $\big\langle\bm{\Pi}\big\rangle$,
  \begin{equation}
  \bm{f}_\mathrm{ac} = -\boldsymbol{\nabla}\cdot\big\langle\bm{\Pi}\big\rangle.
  \end{equation}
The second-order quantity $\big\langle\bm{\Pi}\big\rangle$ is given by products of the first-order acoustic fields $p_1$ and $\bm{v}_1$~\cite{Landau1993},
 \begin{equation}
 \big\langle\bm{\Pi}\big\rangle = \big\langle{p_2}\big\rangle \mathbf{1} + \big\langle{{\rho_0}{\bm{v}_1}{\bm{v}_1}}\big\rangle,
 \end{equation}
where the second-order mean Eulerian excess pressure $\big\langle{p_2}\big\rangle$ takes the form
 \begin{equation}
 \big\langle{p_2}\big\rangle = \frac{1}{4} \kappa_0 |p_1|^2 - \frac{1}{4} \rho_0 |\bm{v}_1|^2.
 \end{equation}
The acoustic body force $\bm{f}_\mathrm{ac}$ was derived on the slow hydrodynamic time scale $\tau$ in Ref.~\cite{Karlsen2016} from the divergence of the time-averaged acoustic momentum-flux-density tensor induced by continuous spatial variations in the fluid density $\rho_0$ and compressibility $\kappa_0$, or equivalently in density $\rho_0$ and sound speed $c_0$,
 \begin{subequations}
 \eqlab{facFull}
 \begin{align}
 \bm{f}_\mathrm{ac} &= - \frac{1}{4} |p_1|^2 \boldsymbol{\nabla}\kappa_0
 - \frac{1}{4} |\bm{v}_1|^2 \bm{\nabla}\rho_0
 \\
 &= \frac14 \big(\kappa_0 |p_1|^2 - \rho_0|\bm{v}_1|^2\big)\frac{\bm{\nabla}\rho_0}{\rho_0}
 + \frac12 \kappa_0 |p_1|^2 \frac{\bm{\nabla}c_0}{c_0}.
 \end{align}
 \end{subequations}

\subsection{Slow-time-scale dynamics}

The dynamics on the slow time scale $\tau$ is governed by the momentum- and mass-continuity equations for the fluid velocity $\vvv(\rrr,\tau)$ and pressure $p(\rrr,\tau)$, and by the advection-diffusion equation for the concentration $s(\rrr,\tau)$ of the solute with diffusivity $D$,~\cite{Karlsen2016}
 \begin{subequations}
 \eqlab{DynamicsSlow}
 \begin{align}
 \eqlab{NSSlow}
 \pp_\tau (\rhoO \vvv) &= \divop \big[ \sigmabf - \rhoO\vvv\vvv \big] + \fffac + \rhoO \gvec , \\
 \eqlab{ContSlow}
 \pp_\tau \rhoO &= - \divop \big( \rhoO \vvv \big) , \\
 \eqlab{DiffusionSlow}
 \pp_\tau s &= - \divop \big[ - D \nablabf s + \vvv s \big] .
 \end{align}
 \end{subequations}
Here, $\gvec$ is the gravitational acceleration, $\sigmabf$ is the fluid stress tensor, and $\fffac$ is the acoustic force density.

\subsection{Boundary-driven acoustic streaming}

The above slow-time-scale velocity field $\vvv$ comprises the acoustic streaming in the general inhomogeneous case, which is the main focus of this work. However, as the inhomogeneity in our system is smeared out by diffusion as time passes, it is helpful to be reminded of the streaming flow in an homogeneous systems. This problem was solved analytically by Lord Rayleigh~\cite{LordRayleigh1884} for an infinite parallel-plate channel of height $H$ with its two plates placed symmetrically around the $x$-$y$ plane at $z = \pm\frac12 H$ and with the imposed first-order standing-wave acoustic fields with wavelength $\lambda$ and wavenumber $k=2\pi/\lambda$ along the $y$ direction: $p_1(y) = p_\mathrm{ac} \sin(ky)\:\ee^{-\ii\omega t}$ and $\bm{v}_1 = v_\mathrm{ac}\cos(ky) \ee^{-\ii\omega t}\:\eee_y$. In the case of  $\lambda \gg H \gg \delta$, where $\delta = \sqrt{2\eta_0/(\rho_0\omega)}$ is the thickness of the viscous boundary layer, Rayleigh found the time-averaged components $\big\langle v_{2y} \big\rangle$ and $\big\langle v_{2z}\big\rangle$ of the second order fluid velocity $\big\langle\bm{v}_2(y,z) \big\rangle$ outside the viscous boundary layer to be
 \begin{subequations}
 \begin{align}
 \eqlab{v2yRayleigh}
 \big\langle v_{2y} \big\rangle &= \frac{3}{8}\frac{v^2_\mathrm{ac}}{c_0}\:\sin(2ky)
 \left[1-3 \frac{(2z)^2}{H^2}\right]\frac12 ,\\
 \eqlab{v2zRayleigh}
 \big\langle v_{2z} \big\rangle &= \frac{3}{8}\frac{v^2_\mathrm{ac}}{c_0}\:\cos(2ky)
 \left[\frac{2z}{H}  - \frac{(2z)^3}{H^3} \right]\:\frac{kH}{2}.
 \end{align}
\end{subequations}
For an analytical solution in a closed rectangular channel, see Ref.~\cite{Muller2013}, the amplitudes $p_\mathrm{ac}$ and $v_\mathrm{ac}$ are related to each other and to the acoustic energy density $E_\mathrm{ac}$ as
 \begin{equation}
 E_\mathrm{ac} = \frac14 \rho_0 v_\mathrm{ac}^2 = \frac14 \kappa_0 p_\mathrm{ac}^2 .
 \end{equation}

\subsection{Numerical simulations of the system}
\seclab{Numerics}

As described in Ref.~\cite{Karlsen2018}, the dynamics in the 2D channel cross-section is solved numerically, under stop-flow conditions with the initial conditions described in \secref{StreamingInhomog}, using a weak-form finite-element implementation in COMSOL Multiphysics~\cite{COMSOL52} with regular rectangular mesh elements. A segregated solver solves~the time-dependent problem in two steps: (i) The fast-timescale acoustics \eqref{FirstOrderEqs} in the inhomogeneous medium is solved while keeping the hydrodynamic degrees of freedom fixed. This allows computation of the time-averaged acoustic force density $\fffac$, \eqref{facFull}. (ii) The slow-timescale dynamics \eqref{DynamicsSlow} is then integrated in time $\tau$ using a generalized alpha solver with a damping parameter of 0.25, and a maximum time step $\Delta \tau=7.5~\SIms$, while keeping the acoustic energy density fixed at $\Eac=50~\SIPa$~\footnote{To fix $\Eac$, which varies due to small shifts in resonance frequency as $s(\rrr,\tau)$ evolves, we compensate by adjusting the sidewall actuation amplitude $d_0(\tau)$ iteratively; see also Ref.~\cite{Muller2012}.}. This implementation extends our previous one limited to iodixanol solutions~\cite{Karlsen2018} by allowing for $\nablabf \cO \neq \zerovec$ and an $s$-dependent diffusivity $D(s)$.

\section{Materials and methods}

\subsection{Experimental setup and materials}
\seclab{SetupMaterials}

The silicon chip consists of a straight channel of length $L = 24$~mm, width $W = 375~\SImum$, and height $H = 133~\SImum$  as sketched in \figref{setup}. The chip was sealed by a pyrex lid of thickness $h_{\mathrm{lid}} = 1$ mm using anodical bonding, and an 18~mm $\times$ 6.4~mm $\times$ 1.0~mm lead zirconate titanate (PZT) transducer (PZT26, Ferroperm Piezoceramics, Denmark) was bonded underneath using cyanoacrylate glue (Loctite Super Glue, Henkel Norden AB, Stockholm, Sweden). At the main channel inlet three streams join in a trifurcation of which the two side streams are routed via a common port while the center stream has a separate port. At the end of the main channel the outlet has the same trifurcated configuration as the inlet. Pieces of silicone tubing (outer diameter 3~mm, inner diameter 1~mm, and length 7~mm) were glued to the chip inlets and outlets. The inlet flow streams were routed via a motorized four port two way diagonal valve so that the flow could be stopped by short-circuit of the side inlets with the center inlet stream, and the outlet stream was routed via 2-port 2-way solenoid valve. The inlets were used for injecting two different liquids, whereas only one outlet was used for collecting the waste, while the other outlet was blocked during all measurements. A PT100 thermo-resistive element was bonded to the PZT transducer to record the temperature.

\begin{figure}[!t]
\centering
\includegraphics[width=0.8\columnwidth]{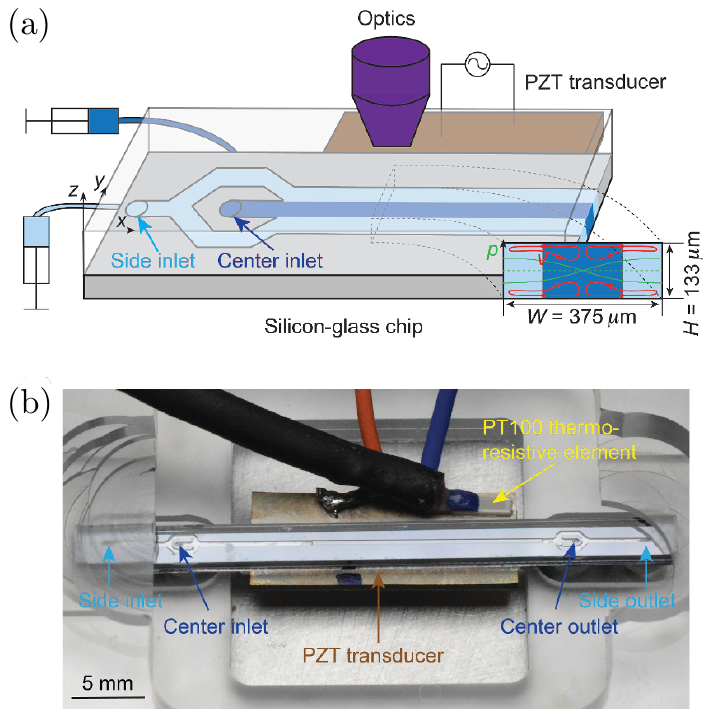}
\caption[]{\figlab{setup}
(a) Sketch of the acoustofluidic silicon chip (gray) sealed with a glass lid, which allows optical recording (purple) of the tracer bead motion (red trajectories) in the channel cross section of width $W = 375 \,\SImum$ and height $H = 133 \,\SImum$. A Ficoll solution (dark blue) is injected in the center and laminated by pure water (light blue). The piezoelectric transducer (brown) excites the resonant half-wave pressure field $p_1$ (inset, green) at 2 MHz. (b) Top-view photograph of the chip (dark gray) mounted on the PZT transducer (brown) and placed in its holder (transparent plastic).}
\end{figure}
\renewcommand\arraystretch{1.1}
\begin{table}[!b]
\caption{\tablab{InjectedSolutions} Specification of the nine inhomogeneous solutions S1 -- S9 used in the experiments with $\hat{\rho}_*$ and $\hat{c}_*$ defined in \eqref{rhoScS}. PBS is phosphate-buffered saline.}
\begin{ruledtabular}
\begin{tabular}{ l r l l r r}
ID & \multicolumn{2}{l}{Center inlet} & Side inlet & $\hat{\rho}_*$ (\%) & $\hat{c}_*$ (\%) \\ \noalign{\smallskip}  \hline
S1 &  5\% & Ficoll PM400 & PBS & 0.96 & 0.00 \\
S2 & 10\% & Ficoll PM70 & 6.38\% iodixanol & 0.00 & 1.92 \\
S3 & 15\% & iodixanol & 10\% Ficoll PM70 & 4.74 & $-$2.13 \\ \hline
S4 & 10\% & Ficoll PM400 & Milli-Q water& 3.51 & 1.69 \\
S5 &  5\% & Ficoll PM400 & Milli-Q water& 1.72 & 0.77 \\
S6 &  1\% & Ficoll PM400 & Milli-Q water& 0.34 & 0.19 \\ \hline
S7 & 10\% & Ficoll PM70  & Milli-Q water& 3.51 & 1.69 \\
S8 &  5\% & Ficoll PM70  & Milli-Q water& 1.71 & 0.79 \\
S9 &  1\% & Ficoll PM70  & Milli-Q water& 0.34 & 0.17  \\
\end{tabular}
\end{ruledtabular}
\end{table}

The PZT transducer was driven by a function generator (AFG3022B, Tektronix, Inc., Beaverton, Oregon, USA), and the waveforms of the applied voltages to the transducer were monitored by an oscilloscope (TDS1002, Tektronix, Inc., Beaverton, Oregon, USA).  The liquids were injected to the channel using syringe pumps (neMESYS, Cetoni GmbH, Korbussen, Germany) with flow rates controlled by a computer interface.

The main density and speed-of-sound modifier used for this study was Ficoll (PM70 and PM400, GE Healthcare Biosciences AB, Uppsala, Sweden), of different average molecular mass (PM70: 70,000 mol wt; PM400: 400,000 mol wt). The Ficoll was dissolved in Milli-Q water to different mass fractions. The density and speed of sound of all the solutions were measured using a density and sound velocity meter (DSA 5000 M, Anton Paar GmbH, Graz, Austria), and the viscosities were measured by a falling ball micro viscometer (MINIVIS II, AMETEK Grabner Instruments, Vienna, Austria). We used the nine different solution combinations listed in \tabref{InjectedSolutions} to create inhomogeneous acoustofluidcs.

The material parameters of the  Ficoll PM70 and Ficoll PM400 solutions used in the experiments at temperature $T = 25~\SICel$ are given in the Supplemental Material\ \footnote{See Supplemental Material at [URL] for details on the fitting leading to $\rho_0(s)$, $c_0(s)$, $\eta_0(s)$, and $D(s)$.}. The resulting fitting expressions for these parameters are listed in \tabref{SolutionParameters}.

\renewcommand\arraystretch{1.1}
\begin{table}[!t]
\caption{\tablab{SolutionParameters}  The measured density $\rho_0(s)$, sound speed  $c_0(s)$, and viscosity $\eta_0(s)$, obtained as described in \secref{SetupMaterials}, as well as diffusivity $D(s)$, see \secref{concentration}, for homogeneous Ficoll-Milli-Q solutions as a function of the solute mass fraction (concentration) $s$ the interval $0 < s < 0.1$. The fits are based on 9 (for $D$ only 3) values of $s$ in that interval.}
\begin{ruledtabular}
\begin{tabular}{@{\hspace*{8mm}}rl@{\hspace*{7mm}}}
\multicolumn{2}{l}{Ficoll PM70}\\
$\rho_0(s) =$ & $(1+0.349\:s)\:996.85\:\SIkgm$ \\
$c_0(s) =$ & $(1+0.167\:s)\:1496.30$~m\:s$^{-1}$\\
$\eta_0(s) =$ & $ \exp(10.82\:s)\:0.893$\; mPa\:s\\
$D(s) = $ & $(1 - 5.51\:s + 23.0\:s^2)\:1.21 \times 10^{-10}\:\SIm^2\:\SIs^{-1}$\\
\hline\noalign{\smallskip}
\multicolumn{2}{l}{Ficoll PM400}\\
$\rho_0(s) =$ & $(1+0.348\: s)\:996.91~\SIkgm$ \\
$c_0(s) =$ & $(1+0.164\: s)\:1496.50$~m\:s$^{-1}$\\
$\eta_0(s) =$ & $\exp(16.20\:s)\:0.893$\:mPa\:s\\
$D(s) = $ & $(1 - 10.3\: s + 56.0\: s^2)\:1.15 \times 10^{-10}~\SIm^2\:\SIs^{-1}$\\
\end{tabular}
\end{ruledtabular}
\end{table}

\subsection{The GDPT setup and method}

Fluorescent green polystyrene beads with a nominal diameter of 0.49 $\SImum$ (Molecular Probes, Thermo Fisher Scientific, Waltham, MA, USA) were suspended in the solutions as tracer particles. The images of the motions of tracer particles in the microchip were recorded using a CMOS camera (ORCA-Flash4.0 V3, Hamamatsu Photonics K.K., Japan) mounted on an epi-fluorescence microscope (BX51WI, Olympus Corporation, Tokyo, Japan). An objective lens with 10x  magnification and 0.3 numerical aperture was used and a cylindrical lens with a focal length of 300 mm was placed between the camera and the objective at a distance of 20 mm in front of the camera. This configuration provided a measurement volume of 1.31 $\times$ 1.52 $\times$ 0.15~mm$^3$. Blue light fluorescent excitation light was provided by a double-wavelength LED unit (pE-300$^{\mathrm{ultra}}$, CoolLED Ltd., UK) with a peak wavelength of 488 nm. A standard fluorescence filter cube was used with an excitation pass-band from 460~nm to 490~nm and a high-pass emission filter at 520 nm.

The motion of the tracer particles was recorded using a general defocusing particle tracking (GDPT) technique~\cite{Barnkob2015, GDPTlab2018}. GDPT is a single-camera particle tracking method in which astigmatic images are employed by using a cylindrical lens. An unique defocused elliptical shape of a spherical particle in the depth coordinate ($z$-coordinate) can be provided in such a system which enables robust three-dimensional tracking of particle motion in microfluidic systems. Before performing the measurement, a stack of calibration images was obtained with an interval of 1 $\SImum$ in depth coordinate by moving a motorized \emph{z}-stage (MFD, M\"{a}rzh\"{a}user, Wetzlar GmbH \& Co. KG, Wetzlar, Germany) equipped on the microscope. Then, the height of the stage was fixed and the motion of the particles was recorded. The image acquisition was performed with an exposure time of 90~ms and a frame rate of 10~fps. The acquired images were analyzed in GDPTlab by performing normalized cross-correlation and comparing the acquired images with the calibration stack. Because the channel is filled with liquid, the values of liquid refractive indices are required for calculating the true particle position in $z$-coordinate, which were measured using an automatic refractometer (Abbemat MW, Anton Paar GmbH, Graz, Austria). The mean value of the refractive indices of the two liquids injected to the channel was used for particle tracking, which gives a maximum error of 1~$\SImum$ in the $z$ direction. Finally, the particle trajectories and velocities were constructed. Particles were rejected if their cross-correlation peak amplitude was less than 0.95 and trajectories were rejected if they had less than six particle positions.

\subsection{Experimental procedures}

A laminated flow of two liquids was injected to the channel to form a concentration gradient with a flow rate of 100 $\SImuL$/min and a volumetric ratio near unity, see \figref{setup}. Before and during the measurements, the transducer was actuated by a linear frequency sweep from 1.95 to 2.05~MHz in cycles of 1~ms to produce a standing half-wave across the width~\cite{Manneberg2009a}. The frequency sweep covers the identified resonance frequencies at 1.96 MHz for pure water and 1.97 MHz for 10\% Ficoll PM400 and ensures steady actuation throughout the experiment during the time-evolution of the concentration field. The applied voltages (ranging from 1.59 to 1.67 V peak-to-peak) were adjusted for each injection of fluids to maintain the same acoustic energy density $E_\mathrm{ac} \approx 52$~Pa in the channel. For the inhomogeneous situation with three liquid layers, we estimated $E_\mathrm{ac} = \frac12(E_\mathrm{ac}^\mathrm{cntr} + E_\mathrm{ac}^\mathrm{side})$, where $E_\mathrm{ac}^\mathrm{cntr}$ and $E_\mathrm{ac}^\mathrm{side}$  in the center and side layers were measured in their respective homogeneous states by tracking individual polystyrene beads with a nominal diameter of 6.33~$\SImum$ (PFP-6052, Kisker Biotech GmbH~\&~Co. KG, Steinfurt, Germany)~\cite{Barnkob2010}. At time $\tau=0$, the flow was stopped, and the images for the GDPT measurements were recorded. The instantaneous stop of the flow in the channel was performed by short-circuit of the two inlets by switching the four-port valve, see \figref{procedure}, which stops the flows and equilibrates the pressures of the two inlet streams. For each set of measurements, the particle motion was recorded for $200~\mathrm{s}$ to observe the evolution of the acoustic streaming. Each measurement was repeated at least 16 times to improve the statistics.

\begin{figure}[t]
\centering
\includegraphics[width=0.75\columnwidth]{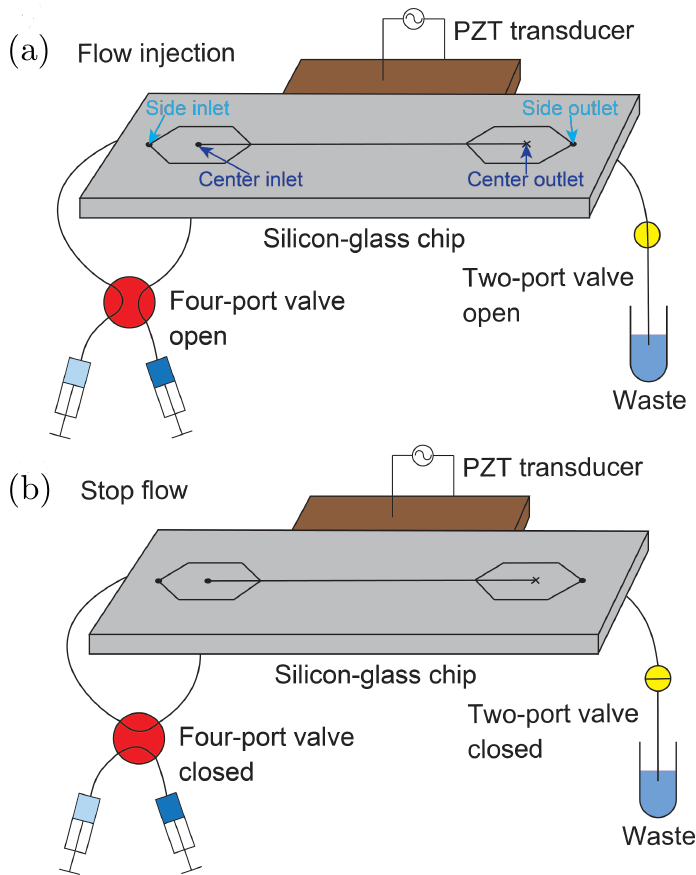}
\caption[]{\figlab{procedure}
Sketch of the stop-flow mechanism. (a) When the two liquids are injected, the two syringes are connected to the two inlets through the open four-port valve, and the waste is collected through the open two-port valve. (b) To stop the flow, the two inlets are short circuited by closing the four-port valve, and the outlet is blocked by closing the two-port valve. The center outlet is always blocked during the experiment.}
\end{figure}

The evolution of the concentration gradient in the channel was mapped by confocal microscopy (Fluoview 300, Olympus Corporation, Tokyo, Japan) in the $x$-$y$ plane at the mid-height of the channel ($z = 0$). The same objective lens as in the streaming measurement was used and a scan rate of 0.89 s$^{\mathrm{-1}}$ was chosen, which provided a measurement area of 658~$\SImum$ $\times$ 385~$\SImum$. To trace the Ficoll concentration fields fluorescein isothiocyanate (FITC)-labeled Ficoll (Polysucrose 70- and Polysucrose 400-fluorescein isothiocyanate conjugate, Sigma-Aldrich Sweden AB, Stockholm, Sweden) were added to the solutions in amounts ranging from 0.10~\% to 0.16~\%. Before the measurement of the concentration gradient a background image was recorded when no fluorescent molecules were present in the channel. A linear decay of the intensity of the fluorescence signal emitted from FITC-labelled Ficoll solutions with the decreasing concentration was confirmed. After exciting the sound field, the two liquids were laminated in the channel by infusing them with a total flow rate of 100~$\SImuL$/min, and it therefore took them $\sim 1$~s to reach the observation region which is 10 mm downstream from the trifurcation inlet. The energy density, the flow rate, and the volumetric ratio were the same as those in the streaming measurements. The image acquisition started at $\tau = 5$~s after stopping the flow, and it continued until $\tau = 195$ s in intervals of 10 s. Each measurement was repeated three times.

\section{Results and discussion}

\begin{figure}[!b]
\centering
\includegraphics[width=0.8\columnwidth]{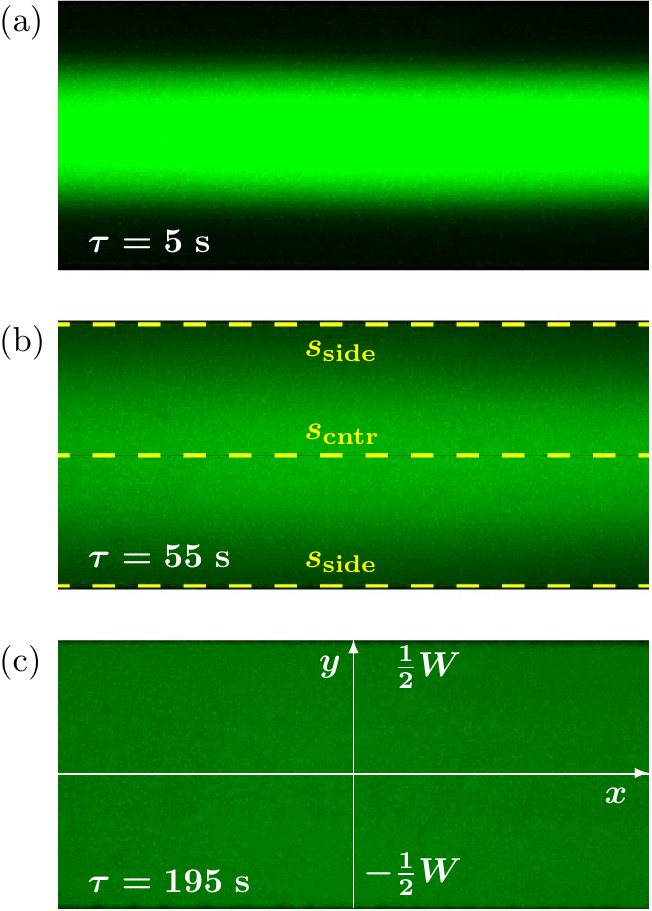}
\caption[]{\figlab{sDevelopment}
Confocal images in the horizontal $x$-$y$ plane taken of solution S4 with acoustics on at (a) $\tau$ = 5~s, (b) $\tau$ = 55~s, and (c) $\tau$ = 195~s. The yellow lines indicate the locations where $s_\mathrm{cntr}$ and $s_\mathrm{side}$ are measured, which are then used to determine $\hat{\rho}_*$, $\hat{c}_*$, and $\hat{s}_*$, see \eqsref{rhoScS}{sStarDef}.}
\end{figure}

\subsection{Time evolution of concentration fields}
\seclab{concentration}

The acoustofluidics of the inhomogeneous system is governed by the time evolution of the molecular concentration field $s$. By adding fluorescently tagged Ficoll molecules to the center flow stream, we studied this evolution by confocal microscopy. For a given solution combination injected, and confocal $x$-$y$ scans were recorded at mid height ($z = 0$). In \figref{sDevelopment} is shown examples of such scans for solution S4 (10\% PM400 and Milli-Q) of \tabref{InjectedSolutions} recorded at $\tau=5$, 55, and 195~s. From the fluorescence intensity, $\rho_0$ and $c_0$ could be measured at different locations in the channel through a calibration curve using known concentrations $s$. The concentration gradient was quantified by measuring the intensity profile across the channel width, see \figref{sVSy}, which shows that the concentration field evolves from a steep box-shaped distribution at early times to a progressively more flat distribution at later times. Since the measurement plane is placed at $z = 0$, away from the top and bottom boundaries, the evolution of the concentration field is governed purely by diffusion at early times and therefore the advection due to the streaming can be neglected. With this assumption the diffusivity $D$ of each solution can be extracted from the concentration profile $s$ at early times by a simple numerical model of molecular diffusion in 1D in the transverse $y$ direction with zero-flux boundary conditions at the walls $y = \pm\frac12 W$. It should be noted that the diffusivity $D$ is measured in the presence of the ultrasound field, which might be different from the situation without ultrasound owing to the barodiffusion~\cite{Landau1993} and the possible interaction between Ficoll molecules and the sound waves. The resulting expression for $D$ as a function of the solute concentration $s$ is listed in \tabref{SolutionParameters}.

\begin{figure}[!t]
\centering
\includegraphics[width=0.8\columnwidth]{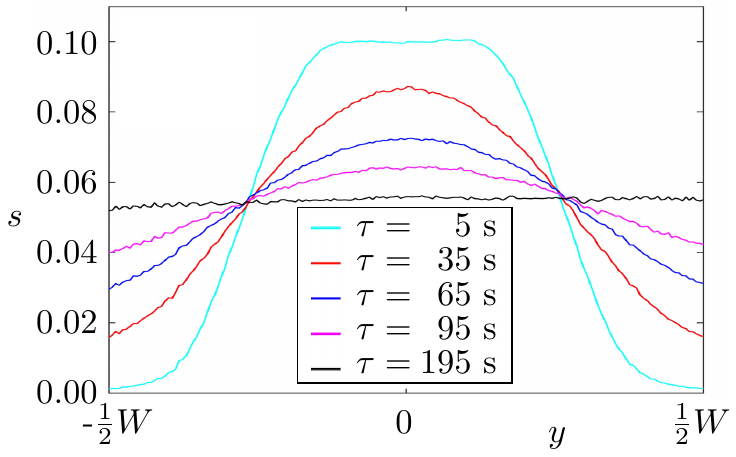}
\caption[]{\figlab{sVSy}
The time evolution with acoustics present of the concentration profile for solution S4 with FITC-labelled Ficoll molecules deduced from recordings as shown in \figref{sDevelopment}.}
\end{figure}

\subsection{Streaming in inhomogeneous solutions}
\seclab{StreamingInhomog}

With the nine different solutions S1--S9 of \tabref{InjectedSolutions} and the general theory summarized in \secref{Theory}, we are now in position for a detailed study of the acoustic streaming in inhomogeneous solutions. We quantify the magnitude of a given inhomogeneity by the excess mass density $\hat{\rho}_*$ and the excess speed of sound $\hat{c}_*$, based on local values in the center and in the side of the channel,  see \figref{sDevelopment}(b),
 \begin{equation} \eqlab{rhoScS}
 \hat{\rho}_* = \frac{\rho_\mathrm{cntr}}{\rho_\mathrm{side}} -1, \qquad
 \hat{c}_* = \frac{c_\mathrm{cntr}}{c_\mathrm{side}} -1.
 \end{equation}

We begin by studying solutions S1--S4 that, as can be seen from \tabref{InjectedSolutions}, are chosen for their specific dependencies of $\hat{\rho}_*$ and $\hat{c}_*$: S1 depends only on $\hat{\rho}_*$, S2 on only $\hat{c}_*$, S3 on both with opposite sign, and S4 on both with the same sign. Moreover, in all four cases, the center liquid was chosen so as to be stabilized by the acoustic body force acting on the inhomogeneous fluid, which avoided undesirable particle motion due to relocation of the two liquids~\cite{Deshmukh2014}. The overlaid particle positions from $\tau = 20$ to 30~s in different gradients for S1--S4 are shown in \figref{FourStreaming}. It is seen that streaming is suppressed in the bulk for all four solutions and only manifested by flat streaming rolls located near the top and bottom walls at $z = \pm \frac12 H$, in full agreement with our previous findings for an iodixanol solution with only density dependency and no sound speed dependency \cite{Karlsen2018}. All streaming patterns are similar, exhibiting four vortices with no apparent symmetry around the vortex centers and having a larger width close to the center $y=0$ than to the side walls at $y=\pm\frac12 W$.

\begin{figure}[!t]
\centering
\includegraphics[width=0.9\columnwidth]{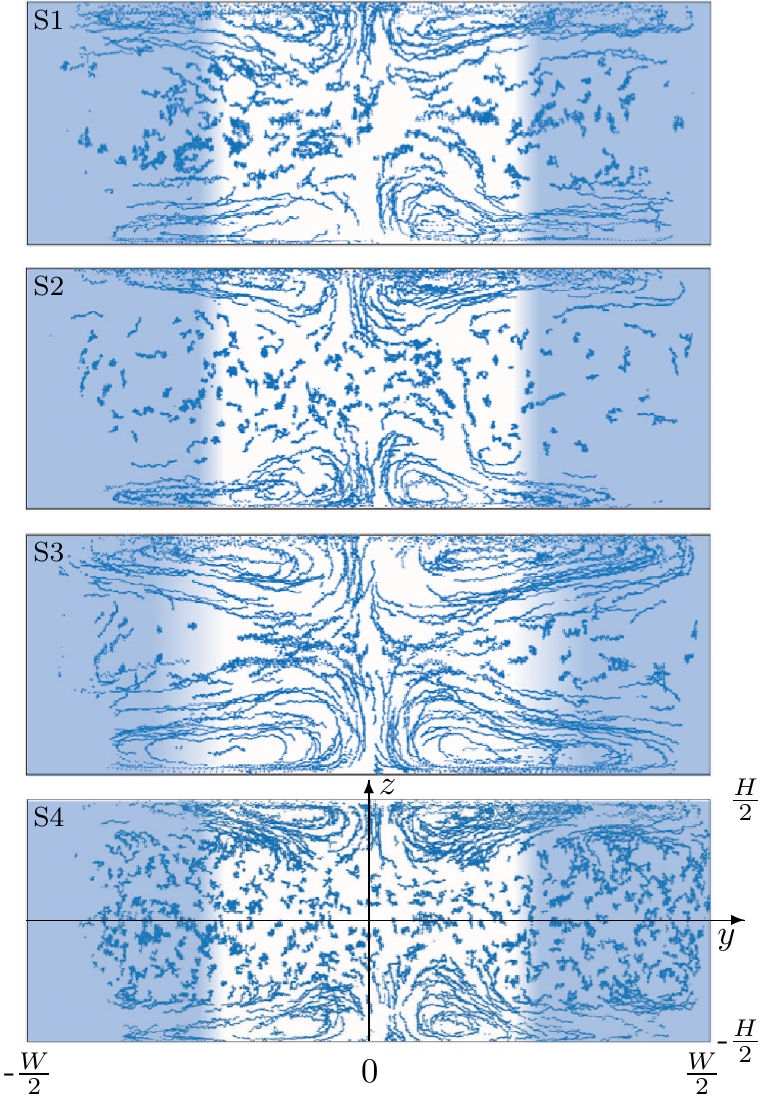}
\caption[]{\figlab{FourStreaming}
The particle positions (blue points) in the vertical $y$-$z$ cross section of width $W=375~\SImum$ and height $H=133~\SImum$ overlaid from 100 frames between $\tau$ = 20~s and 30~s for the inhomogeneous solutions S1, S2, S3, and S4 listed in \tabref{InjectedSolutions}. The color plot represents the concentration of the solute molecules from low (dark) to high (white).}
\end{figure}

\begin{figure}[!t]
\centering
\includegraphics[width=0.9\columnwidth]{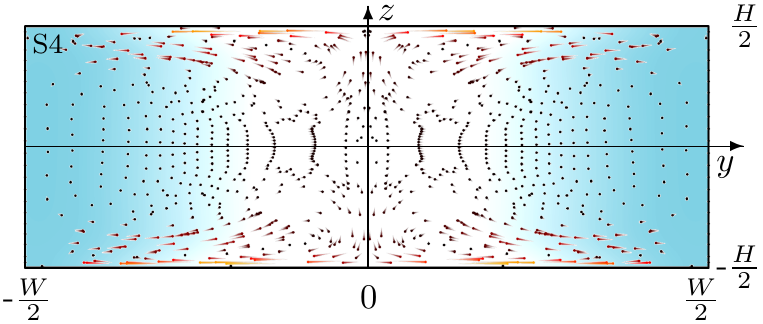}
\caption[]{\figlab{ComsolStreaming}
Numerical simulation (see \secref{Numerics}) in the vertical $y$-$z$ cross section for solution S4 (10\% Ficoll PM400 and Milli-Q) of \tabref{InjectedSolutions} with the parameters given in \tabref{SolutionParameters} in a standing half-wave pressure field of energy density $\Eac = 50$~Pa corresponding to S4 in \figref{FourStreaming}. The comet-tail plot shows the position (dots) and velocity (colored comet tails) of 1000 polystyrene 500-nm-diameter tracer particles at $\tau = 25$~s that started out in a regular 50$\times$20 mesh at $\tau = 0$~s. The color plot represents the concentration of the solute molecules from $s\approx 0.015$ (blue) to $s\approx 0.085$ (white).}
\end{figure}

The asymmetry in the streaming rolls can be explained by the evolution of the concentration field $s(\rrr,\tau)$ near the top and the bottom walls at early times. Initially by construction, $s$ exhibits a steep, nearly vertical box-like distribution and the acoustic body force $\bm{f}_\mathrm{ac}$ stabilizes the fluid in the center stream and prevents advective recirculation in the bulk~\cite{Karlsen2017}. However, near the top and bottom walls, the compressed streaming flow transports the center solution towards the sides, causing wedges with nearly $45^\circ$ slopes to form in the concentration field $s$ at regions near $y=0$ and $z=\pm \frac12 H$. Because $\bm{f}_\mathrm{ac}$ is parallel to the concentration gradient, the wedges led to a weaker $\bm{f}_\mathrm{ac}$ in the horizontal direction and therefore the streaming rolls have larger curvature near the center of the channel. These wedges are difficult to resolve experimentally, so to confirm the above hypothesis, we performed a numerical simulation of the evolution of the concentration field and the streaming field using COMSOL Multiphysics as described in \secref{Numerics}. In \figref{ComsolStreaming} we show the results of such a simulation for solution S4 at time $\tau=25$~s after injection in the device, in which a transverse standing half-wave is present as in the experiment. The wedge shape in the concentration field is clearly seen in the transition form high (white) to low (blue) concentration near the top center and bottom center of the cross section. Moreover, the simulated particle motion show the observed asymmetric vortices being broader near the center $y \approx 0$ compared to the sides at $y \approx \pm\frac12 W$.

In the Supplemental Material\ \footnote{See Supplemental Material at [URL] for COMSOL simulations of the time evolution of the concentration field $s$ and the trajectories of 500-nm-diameter and 1000-nm-diameter tracer particles in solution S4.} we have placed movies showing the time evolution for $0 < \tau < 40$~s of which \figref{ComsolStreaming} is the single frame at $\tau = 25$~s.

\subsection{Time evolution of the streaming suppression}

To follow the time evolution and final breakdown of the streaming suppression, the tracer particle motion was tracked for 200 s after stopping the flow. The streaming evolution is shown in \figref{StreamingDevelop} for the inhomogeneous Ficoll-water solution S7 of \tabref{InjectedSolutions}. At early times, $\tau = 35$\:s in \figref{StreamingDevelop}, the streaming is greatly suppressed in the bulk and the four streaming rolls are confined to the boundaries with an asymmetric pattern. As time evolves, $\tau = 105$\:s in \figref{StreamingDevelop}, the streaming rolls grow towards homogeneous steady state Rayleigh streaming, but the asymmetric pattern is still apparent. At later times, $\tau = 195$\:s in \figref{StreamingDevelop}, the streaming pattern is identical to that of an homogeneous system as diffusion has homogenized the system.

To quantify the suppression of streaming, we use the streaming vortex size $\Delta$ that we introduced in Ref.~\cite{Karlsen2018}, defined as the distance between the center of the flow roll, situated at $y = \pm\frac{1}{4}W$ where streaming velocity is zero, and the nearest wall. In all homogeneous states, we find $\Delta_\mathrm{hom} =  (27.4 \pm 2.1)~\SImum$ close to $\big(\frac12 - \frac{1}{\sqrt{12}}\big)H = 28.1~\SImum$ found from $\avr{v_{2y}}=0$ in \eqref{v2yRayleigh}. We then study the time evolution of the six inhomogeneous solutions S4 -- S9 of \tabref{InjectedSolutions}, all created by injecting a given Ficoll solution into the center inlet and milli-Q water into the side inlets. The time evolution of the streaming flow is characterized by the normalized vortex size $\Delta/\Delta_\mathrm{hom}$, as shown in \figref{DeltaTau}(a) for Ficol PM70. We see that $\Delta/\Delta_\mathrm{hom}$ increases slowly at early times and then undergoes a transition to a faster increase. The transition occurs at different times for different Ficoll concentrations. This indicates that the evolution of the concentration field is dominated by diffusion at early times, whereas the advection  due to the streaming plays a minor role. When reaching a critically

\begin{widetext}
\mbox{}\noindent
\begin{figure}[h!]
\centering
\includegraphics[width=1.0\textwidth]{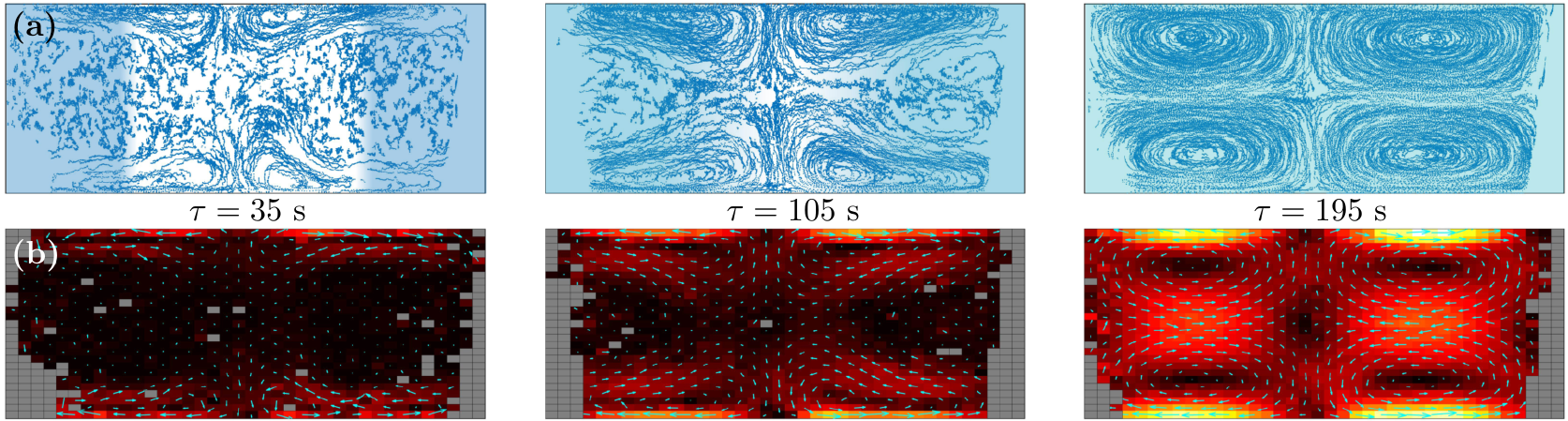}
\caption[]{\figlab{StreamingDevelop}
The acoustic streaming observed in the vertical $y$-$z$ cross section of width $W=375~\SImum$  and height $H=133~\SImum$ at time $\tau$ = 5, 35, and 195~s using the 10~\% Ficoll PM70 solution S7 of \tabref{InjectedSolutions}. (a) Experimental particle positions (blue points) with a color plot of the solute concentration as in \figref{FourStreaming}. (b) Color plot of the streaming velocity amplitude $\left|\avr{\vvvII}\right|$ from $0~\SImum/\SIs$ (black) to $45~\SImum/\SIs$ (white) overlaid with a vector plot (cyan) of $\avr{\vvvII}$. Spatial bins with no data points are excluded (gray).}
\end{figure}
\end{widetext}

\noindent   weak inhomogeneity, the streaming rolls have grown sufficiently so that advection starts to play a more important role. A transition occurs, after which the rate of change of $\Delta/\Delta_\mathrm{hom}$ is increased, since the inhomogeneity now is weakened by both diffusion and advection. This transition occurs at earlier times if the initial concentration of Ficoll is lower for two reasons: The initial solution gradients are weaker and the diffusivity is larger (see \tabref{SolutionParameters}), resulting in a weaker $\bm{f}_\mathrm{ac}$. Hence, at a given time $\tau$, the transition occurs earlier and the rate of change of $\Delta/\Delta_\mathrm{hom}$ is larger for a lower initial Ficoll concentration compared to a higher one. For all concentrations, the streaming rolls expand from the walls into the bulk as the inhomogeneity is smeared out, and finally they become the same as for homogeneous streaming, indicated by the same level of $\Delta/\Delta_\mathrm{hom}$ at late times in \figref{DeltaTau}(a).

\begin{figure}[!t]
\centering
\includegraphics[width=0.95\columnwidth]{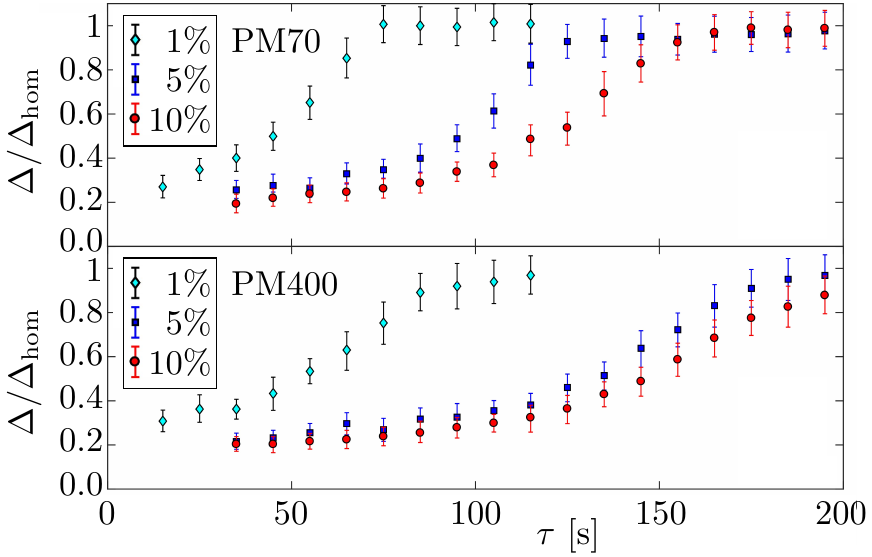}
\caption[]{\figlab{DeltaTau}
The normalized vortex size $\Delta/\Delta_\mathrm{hom}$ versus time $\tau$ using (a) Ficoll PM70 solution S7 -- S9 and (b) Ficoll PM400 solution S4 -- S6 with 1\%, 5\%, or 10\% mass concentration in the center inlet and Milli-Q water in the side inlets. $\Delta(\tau)$ is calculated from overlaid 100 frames recorded in intervals of 0.1~s from $\tau - 5$~s to $\tau + 5$~s, so each data point represents a time interval of 10~s. Each error bar is the error computed when fitting the raw data by a quadratic function in $y$ to determine the center of the streaming vortex.}
\end{figure}

The time evolution also depends on the mass of the molecules that cause the inhomogeneity, which corresponds to the diffusivity.  In \figref{DeltaTau}(b) is shown the streaming roll evolution for Ficoll PM400, which has 5.7 times larger mass and 25~\% lower diffusivity (at $s = 0.1$) compared to Ficoll PM70 in  \figref{DeltaTau}(a). The rate of change of $\Delta/\Delta_\mathrm{hom}$ is lower and the transition point between diffusion and advection-diffusion dominated regimes are shifted to later times for all initial concentrations.

To further validate that the evolution of $\Delta$ is dominated by diffusion at early times, we plot in \figref{DeltaTauRescaled} $\Delta/\Delta_\mathrm{hom}$ versus the rescaled time $\tau/\tau_\mr{diff}$, where $\tau_\mr{diff} = (\frac14 W)^2/(2D)$ is the diffusion time for the given solute. By this rescaling, the difference in diffusivity between the solutions is removed, and a nearly perfect collapse of the six data sets is observed for $\tau \lesssim 2\tau_\mr{diff}$. For $\tau \approx 2\tau_\mr{diff}$ the previously described transition to the advection-diffusion regime occurs, and for $\tau \gtrsim2\tau_\mr{diff}$ the collapse is not as good, as the advection part does not scale with the diffusion time. In \figref{DeltaTauRescaled} we see a higher rate of change and a larger spread in the data points after the transition. As $\fffac$ is weak in the two 1\% solutions, advection plays a role from the beginning. Time zero is therefore ill-defined, and consequently, we have shifted these two data sets by $0.9\:\tau_\mr{diff}$ in time to make the transition point coincide with that of the four 5~\% and 10~\% solutions.

\begin{figure}[!t]
\centering
\includegraphics[width=0.9\columnwidth]{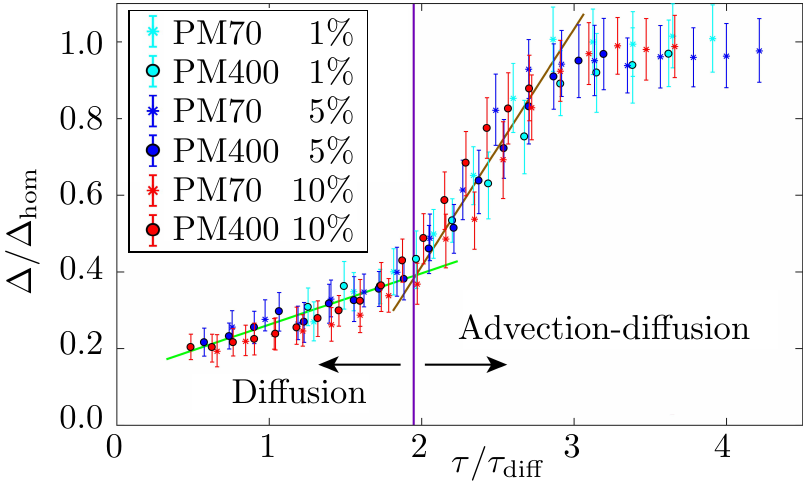}
\caption[]{\figlab{DeltaTauRescaled}
Plot of the normalized vortex size $\Delta/\Delta_\mathrm{hom}$ versus rescaled time $\tau/\tau_\mathrm{diff}$, where $\tau_\mathrm{diff} = (\frac14 W)^2/(2D)$ is the diffusion time for the given solute molecule, for Ficoll solutions S4--S9 from \figref{DeltaTau}. The green line indicates the early time diffusion-dominated dynamics, while the brown line indicate the late time advection-diffusion-dominated dynamics. The two 1\% solutions have been shifted $0.9\:\tau_\mr{diff}$, see the text.}
\end{figure}

\begin{figure}[!b]
\centering
\includegraphics[width=0.9\columnwidth]{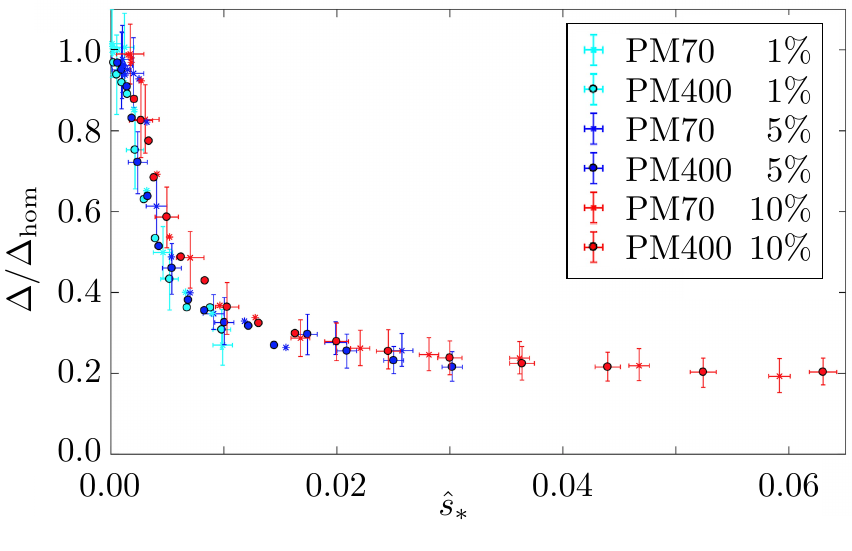}
\caption[]{\figlab{DeltaSstar}
The normalized vortex size $\Delta/\Delta_\mathrm{hom}$ plotted versus the normalized concentration difference $\hat{s}_*$ for the six different Ficoll solutions S4--S9. For clarity, the error bars are only shown for every third data point for $\hat{s}_* < 0.02$.}
\end{figure}

In a final analysis, we tie the evolution of the normalized vortex size $\Delta/\Delta_\mathrm{hom}$ directly with the underlying concentration difference between the center and the sides of the sample. By a first-order Taylor expansion of the inhomogeneous density $\rhoO(s)$ and sound speed $\cO(s)$, we define the normalized concentration difference $\hat{s}_*$ as
 \begin{equation} \eqlab{sStarDef}
 \hat{s}_* = \hat{\rho}_* \frac{\rhoO(0)}{\frac{\dm}{\dm s}\:\rhoO(0)} = \hat{c}_* \frac{\cO(0)}{\frac{\dm}{\dm s}\:\cO(0)}.
 \end{equation}
In \figref{DeltaSstar} we plot $\Delta/\Delta_\mathrm{hom}$ for all six Ficoll-Milli-Q solutions S4--S9 of \tabref{InjectedSolutions} as a function of $\hat{s}_*$. We observe that all the six data sets fall on a single curve that increases as $s$ decreases, giving strong support to the hypothesis that the suppression of the acoustic streaming in the bulk is governed by $\bm{f}_\mathrm{ac}$ resulting from the concentration profile $s$.

\section{Conclusions}

In this paper, we investigated experimentally acoustic streaming, in a half wavelength resonator, for aqueous solutions that was made spatially inhomogeneous in density and compressibility by a solute concentration gradient. The results show that acoustic streaming patterns are very sensitive to such inhomogeneities. Acoustic streaming in the bulk of inhomogeneous fluids is suppressed by confinement of the recirculating streaming rolls near the boundaries parallel to the direction of sound propagation and the suppression is caused by an inhomogeneity-induced acoustic body force $\bm{f}_\mathrm{ac}$. The suppressed streaming rolls exhibit an asymmetry pattern due to a local streaming-induced deformation of the molecular concentration field near the walls where streaming is generated. The streaming rolls grow over time due primarily to diffusion, but for late times advection play an important role as $\bm{f}_\mathrm{ac}$ vanishes and the system becomes homogeneous. For Ficoll solutions, $\bm{f}_\mathrm{ac}$ decays steadily over a time span of 100 s to 200 s and a significant shift in $\Delta$ was detected for inhomogeneities in the solute mass fraction down to $0.1~\%$. We see a clear potential for this type of acoustic streaming suppression to enable acoustic manipulation, enrichment, and fractionation of particles in the sub-micrometer range by acoustophoresis.

\section{Acknowledgements}

W. Qiu was supported by the People Programme (Marie Curie Actions) EC-FP7/2007-2013,
REA Grant No.~609405 (COFUNDPostdocDTU). P. Augustsson was supported by Starting Grant no. 2016-04836 from the Swedish Research Council. We are grateful to R. Barnkob and M. Rossi, Universität der Bundeswehr München, for providing the software GDPTlab~\cite{Barnkob2015, GDPTlab2018}, and Jeppe Revall Frisvad, Technical University of Denmark, for accessing refractometer.\\

%%%%%%%%%%%%%%%%%%%%%%%%%%%%%%%%%%%%%%%%%%%%%%%%%%%%%%%%%%%%%%%%%%%
%
% Bibliography
%
%%%%%%%%%%%%%%%%%%%%%%%%%%%%%%%%%%%%%%%%%%%%%%%%%%%%%%%%%%%%%%%%%%%

%\bibliographystyle{apsrev4-1-titles}
%\bibliography{acoustofluidics}


\end{document}